\documentclass[preprint,prd,nofootinbib,tightenlines,amsmath]{revtex4}
\usepackage{enumerate}
\usepackage{multirow}

\usepackage{epsfig}
\usepackage{graphics,color}      % usual driver
\usepackage{verbatim}
\usepackage{amsmath}    % need for subequations
\catcode`\@=11
\def\sla@#1#2#3#4#5{{%
 \setbox\z@\hbox{$\m@th#4#5$}%
 \setbox\tw@\hbox{$\m@th#4#1$}%
 \dimen4\wd\ifdim\wd\z@<\wd\tw@\tw@\else\z@\fi
 \dimen@\ht\tw@
 \advance\dimen@-\dp\tw@ \advance\dimen@-\ht\z@
 \advance\dimen@\dp\z@
 \divide\dimen@\tw@ \advance\dimen@-#3\ht\tw@
 \advance\dimen@-#3\dp\tw@ \dimen@ii#2\wd\z@
 \raise-\dimen@\hbox to\dimen4{%
 \hss\kern\dimen@ii\box\tw@\kern-\dimen@ii\hss}%
 \llap{\hbox to\dimen4{\hss\box\z@\hss}}}}

\def\slashed#1{%
 \expandafter\ifx\csname sla@\string#1\endcsname\relax
{\mathpalette{\sla@/00}{#1}}
\fi}
\def\declareslashed#1#2#3#4#5{%
 \expandafter\def\csname sla@\string#5\endcsname{%
#1{\mathpalette{\sla@{#2}{#3}{#4}}{#5}}}}
 \catcode`\@=12
\declareslashed{}{/}{.08}{0}{D}
 \declareslashed{}{/}{.1}{0}{A}
 \declareslashed{}{/}{0}{-.05}{k}
 \declareslashed{}{/}{.1}{0}{\partial}
 \declareslashed{}{\not}{-.6}{0}{f}

\def\lsim{\mathrel {\vcenter {\baselineskip 0pt \kern 0pt
    \hbox{$<$} \kern 0pt \hbox{$\sim$} }}}
\def\gsim{\mathrel {\vcenter {\baselineskip 0pt \kern 0pt
    \hbox{$>$} \kern 0pt \hbox{$\sim$} }}}

\newcommand{\bea}{\begin{eqnarray}}
\newcommand{\eea}{\end{eqnarray}}

\begin{document}

\baselineskip=15pt
\preprint{}

\title{Constraining $\tau$-lepton dipole moments and gluon couplings at the LHC}

\author{Alper Hayreter and German Valencia}

\email{valencia@iastate.edu}

\affiliation{Department of Physics, Iowa State University, Ames, IA 50011.}

\date{\today}

\vskip 1cm
\begin{abstract}
We study the constraints that can be placed on anomalous $\tau$-lepton couplings at the LHC. We use an effective Lagrangian description for physics beyond the standard model which contains the $\tau$-lepton anomalous magnetic moment, electric dipole moment and weak dipole moments in two operators of dimension six. We include in our study two additional operators of dimension eight that directly couple the $\tau$-leptons to gluons and are therefore enhanced at the LHC. We consider the two main effects from these couplings: modifications to the Drell-Yan cross-section and  to the $\tau$-lepton pair production in association with a Higgs boson. We find that a measurement of the former at the 14\% level can produce constraints comparable to existing ones for the anomalous dipole couplings; and that a bound on the latter at a sensitivity level of $500\ \sigma_{SM}$ or better would produce the best constraint on the $\tau$-gluonic couplings.

\end{abstract}

\pacs{PACS numbers: }

\maketitle

\section{Introduction}

The study of $\tau$-leptons by the ATLAS and CMS collaborations has matured to the point where they have a very active program. Among the many studies, we can cite examples of standard model (SM) physics with the reconstruction of the $Z$-resonance in the di-tau mode \cite{Chatrchyan:2011nv,Aad:2011kt}; beyond standard model physics (BSM) where heavy resonances such as $Z^\prime$ bosons have been excluded in this channel up to masses around 1~TeV \cite{Chatrchyan:2012hd,Aad:2012gm}. Of course there is also an active program to observe the recently discovered 126~GeV state in the di-tau channel in order to confirm it as the SM Higgs boson \cite{Aad:2012mea,Chatrchyan:2012vp}, as well as searches for additional neutral scalars that decay to $\tau^+\tau^-$.

In order to study BSM physics in a model independent way we can use the effective Lagrangian and the complete catalog of operators up to dimension six that exists in the literature \cite{Buchmuller:1985jz,Grzadkowski:2010es}. In this paper we will focus our attention on the $\tau$-lepton dipole-type couplings: its anomalous magnetic moment and electric dipole moment given by $a_\tau^\gamma$ and $d_\tau^\gamma$ respectively,
\begin{eqnarray}
{\cal L}=\frac{e}{2}\ \bar{\ell}\ \sigma^{\mu\nu}\left(a_\ell^\gamma+i\gamma_5 d_\ell^\gamma \right) \ \ell \ F_{\mu\nu};
\label{defedm}
\end{eqnarray}
and its corresponding weak dipole moments $a_\tau^Z$ and $d_\tau^Z$,
\begin{eqnarray}
{\cal L}=\frac{g}{2\cos\theta_W}\ \bar{\ell}\ \sigma^{\mu\nu}\left(a_\ell^Z+i\gamma_5 d_\ell^Z \right) \ \ell \ Z_{\mu\nu}.
\label{defzdm}
\end{eqnarray}

These usual definitions, Eq.~\ref{defedm}~and~Eq.~\ref{defzdm}, can be generalized to operators that respect the symmetries of the SM
\begin{eqnarray}
{\cal L} = g\frac{d_{\ell W}}{\Lambda^2}\ \bar{\ell}\sigma^{\mu\nu}\frac{\tau^i}{2} e\  \phi W^i_{\mu\nu} + g^\prime\frac{d_{\ell B}}{\Lambda^2}\ \bar{\ell}\sigma^{\mu\nu}e  \ \phi B_{\mu\nu} \  +\ {\rm h.c.}
\label{ginvedm}
\end{eqnarray}
The fully gauge invariant operators contain, amongst other terms, the anomalous magnetic moment, the electric dipole moment (EDM) and the weak dipole moment (ZEDM) of the leptons with the correspondence
\begin{eqnarray}
a_\ell^\gamma &=& \frac{\sqrt{2}\ v }{\Lambda^2}{\rm Re}\left(d_{\ell B}-\frac{d_{\ell W}}{2}\right),\nonumber \\
d_\ell^\gamma &=& \frac{\sqrt{2}\ v }{\Lambda^2}{\rm Im}\left(d_{\ell B}-\frac{d_{\ell W}}{2}\right),\nonumber \\
a_\ell^Z &=& -\frac{\sqrt{2}\ v }{ \Lambda^2}{\rm Re}\left(d_{\ell B}\sin^2\theta_W+\frac{d_{\ell W}}{2}\cos^2\theta_W\right),\nonumber \\
d_\ell^Z &=& -\frac{\sqrt{2}\ v }{ \Lambda^2}{\rm Im}\left(d_{\ell B}\sin^2\theta_W+\frac{d_{\ell W}}{2}\cos^2\theta_W\right).
\label{convertunits}
\end{eqnarray}
where $v$ is the Higgs vacuum expectation value $v\sim 246$~GeV and $\Lambda$ is the scale of new physics. The gauge invariant operators in Eq.~\ref{ginvedm} also relate the anomalous dipole-type couplings to new couplings amongst the $\tau$ lepton, the respective gauge boson and the Higgs boson. Effectively, Eqs.~\ref{defedm}~and~\ref{defzdm} get multiplied by a factor of $(1+h/v)$.

In general BSM operators in the effective Lagrangian with dimension larger than six are further suppressed at low energy by additional powers of the new physics scale. However, for LHC physics, this power counting is altered for dimension eight operators that couple a lepton pair directly to gluons due to the larger parton luminosities \cite{Potter:2012yv}. This motivates us to include in our study the so called "lepton-gluonic" couplings for the $\tau$ which take the form
\begin{eqnarray}
{\cal L} = \frac{g_s^2}{\Lambda^4}\left(d_{\tau G} \ G^{A\mu\nu}G^A_{\mu\nu} \bar \ell_L \ell_R \phi  +d_{\tau \tilde{G}} \ G^{A\mu\nu} \tilde G^A_{\mu\nu} \bar \ell_L \ell_R \phi \right)\ + {\rm h.~c.}
\label{taugluon}
\end{eqnarray}
Here $G^A_{\mu\nu}$ is the gluon field strength tensor and $\tilde G^{A\mu\nu} = (1/2)\epsilon^{\mu\nu\alpha\beta}G^A_{\alpha\beta}$ its dual. 
If we allow for CP violating phases in the coefficients, $d_{\tau G}$ and $d_{\tau \tilde{G}}$, the resulting gluon-lepton couplings take the form
\begin{eqnarray}
{\cal L} &=& \frac{v}{\sqrt{2}}\frac{ g_s^2}{\Lambda^4}\left({\rm Re}(d_{\tau G}) \ G^{A\mu\nu}G^A_{\mu\nu}  +{\rm Re}(d_{\tau \tilde{G}}) \ G^{A\mu\nu} \tilde G^A_{\mu\nu}\right) \bar \ell \ell \nonumber \\
&+& i\frac{v}{\sqrt{2}}  \frac{ g_s^2}{\Lambda^4}\left({\rm Im}(d_{\tau G}) \ G^{A\mu\nu}G^A_{\mu\nu}  +{\rm Im}(d_{\tau \tilde{G}}) \ G^{A\mu\nu} \tilde G^A_{\mu\nu}\right) \bar \ell \gamma_5 \ell. 
\label{couplings}
\end{eqnarray}
The related couplings that include the Higgs boson are obtained by multiplying Eq.~\ref{couplings} with the factor $(1+h/v)$ as before.

The anomalous couplings in Eqs.~\ref{ginvedm}~and~\ref{taugluon} (as well as the leptons $\ell$) in these equations are understood to carry a generation index, and are in principle different for electrons, muons or $\tau$-leptons. They could be lepton number violating as well, but in this paper we restrict our attention to the case of $\tau$-leptons.  The study of the $\tau$-lepton dipole moments has a long history that includes constraints from both cross-sections and $T$-odd asymmetries \cite{Donoghue:1977bw, Silverman:1982ft, Barr:1988mc,delAguila:1990jg,Goozovat:1991nu,delAguila:1991rm,Bernreuther:1993nd,Cornet:1995pw,Vidal:1998jc,GonzalezSprinberg:2000mk,Bernabeu:2004ww,Bernabeu:2008ii}. At the LHC, we find that the cross-sections are the more interesting observables because the $T$-odd asymmetries are suppressed by the mass of the $\tau$.
 
\section{Drell-Yan production of $\tau$-lepton pairs} 
 
We begin by considering the cross-section for Drell-Yan production of $\tau$-lepton pairs. The new couplings will modify this cross-section and can be probed in one of two ways: by comparing the measured cross-section with its value in the SM, or by looking for deviations from lepton universality in measured cross-sections. With the aid of  {\tt MadGraph5} \cite{MadGraph} and {\tt FeynRules} \cite{Christensen:2008py} we compute the contribution of the new couplings to $pp\to \tau^+\tau^-$ at 14 TeV. We distinguish between two different regions of $m_{\tau\tau}$ invariant mass: the high energy region and the $Z$-resonance region. Our numerical results for sample values of the new couplings, as well as fits to these MonteCarlo (MC) points that can be used to interpolate between them are presented in the Appendix.

Our first study corresponds to the high energy region, defined by $m_{\tau\tau}>120$~GeV. This is the most interesting region at LHC for several reasons: it probes phase space not covered by LEP or LEP2; it excludes the $Z$-resonance and therefore significantly reducing the SM background.  The results of our numerical simulations are displayed in Figure~\ref{dtvsig14} and the fits given in Eq.~\ref{fitV} for the electric and weak dipole moments and in Figure~\ref{dtgsig14} and  Eq.~\ref{fitg} for the lepton-gluonic couplings. Our resulting cross-sections are approximately quadratic in the anomalous couplings indicating that the interference with the SM is small. An analytic calculation reveals that the interference terms between the SM and the dipole-type couplings in Eq.~\ref{fitV}  are small due to a suppression by the $\tau$-mass.  We have checked that these interference terms do scale linearly with $m_\tau$ in our MC simulation. For the $\tau$-gluonic couplings there is of-course no interference with the 
SM as the two contributions arise from different parton level processes. 

\begin{figure}[thb]
\includegraphics[width=0.46\textwidth]{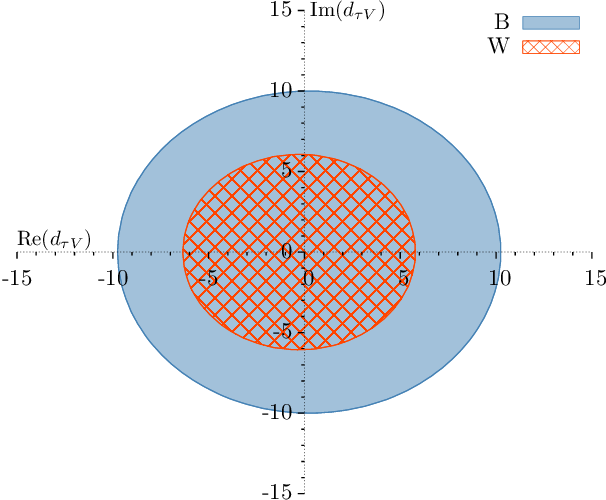} \hspace{0.5cm}
\includegraphics[width=0.45\textwidth]{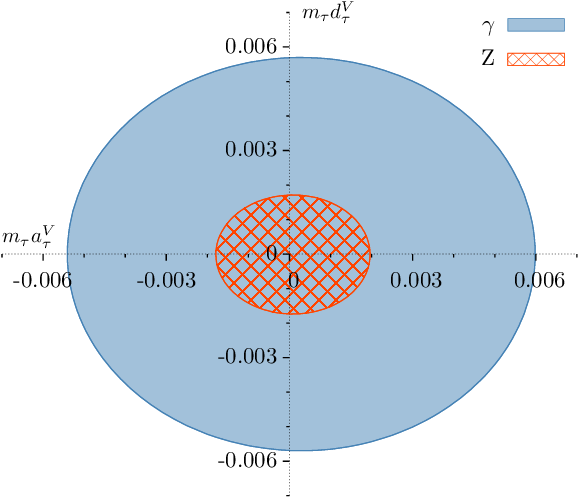}
\caption{Parameter space allowed at $1\sigma$ by requiring the total cross-section $pp\to \tau^+\tau^-$ at 14~TeV with $m_{\tau\tau}>120$~GeV to deviate from its SM value by at most $14\%$}
\label{contourhigh}
\end{figure}

The Drell-Yan cross-section for di-muons and di-electrons above the $Z$-resonance has been measured at the LHC \cite{Lanyov:2007ws} so we can expect that in the future it will be possible to compare it to the di-tau mode. We would expect the largest uncertainty in such a comparison to occur in the di-tau channel so we turn to the CMS search for heavy resonances in this mode for guidance. In Ref.~\cite{Chatrchyan:2012hd} it is estimated that the main systematic uncertainty in di-tau events with high invariant mass, $m_{\tau\tau} > 300$~GeV, results from the estimation of background and is in the range $6-14\%$. To estimate the sensitivity of the LHC to the anomalous couplings we will assume that a comparison between the di-tau and the other di-lepton modes at the $14\%$ level will be possible. To this effect, in Figure~\ref{contourhigh}, we use Eq.~\ref{fitV} to illustrate the region of parameter space allowed at $1\sigma$ by requiring the total di-tau cross-section to deviate from its SM  value (equivalently 
from the di-muon cross-section) by at most $14\%$. 

The best bounds from Figure~\ref{contourhigh}, taken one coupling at a time and converting to the $\gamma,Z$ basis are given by
\begin{eqnarray}
|m_\tau d^\gamma_\tau|  \lsim 0.0057, &&
|m_\tau d^Z_\tau| \lsim 0.0017, \nonumber \\
-0.0054 \lsim m_\tau a^\gamma_\tau  \lsim 0.0060, &&
-0.0018\lsim m_\tau a^Z_\tau \lsim 0.0020
\label{ditauresults}
\end{eqnarray}

The best existing limit for the $\tau$-lepton anomalous magnetic moment $(g_\tau-2)/2 = 2\ m_\tau\  a_\tau^\gamma$ comes from the process $e^+e^- \to e^+e^-\tau^+\tau^-$ as measured by Delphi at LEP2 \cite{Abdallah:2003xd}. At the 95\% c.l. it is
\begin{eqnarray}
-0.052 <(2m_\tau a_\tau^\gamma) < 0.013, &{\rm ~or~}& -0.026 < m_\tau a_\tau^\gamma < 0.007
\end{eqnarray}
The best existing limit for the tau-lepton electric dipole moment comes from the study of $e^+e^- \to \tau^+\tau^-$ by Belle \cite{Inami:2002ah} that constrains both the real and imaginary parts of $d_\tau(0)$, where ${\rm Re}(d_\tau(0)) \equiv -ed_\tau^\gamma$. Again, at the 95\% c. l. it is
\begin{eqnarray}
-2.2 \leq {\rm Re}(d_\tau (0)) \leq 4.5 \times 10^{-17}{\rm ~e~cm}, &{\rm ~or~}& -0.002< m_\tau d_\tau^\gamma < 0.0041
\end{eqnarray}
Belle also quotes a bound for an absorptive part of the form factor (which we do not consider here) $-2.5 \leq {\rm Im}(d_\tau (0)) \leq 0.8 \times 10^{-17}{\rm ~e~cm}$. 
As far as the weak dipole moments, Aleph has obtained the best existing bound from studying $e^+e^-\to \tau^+\tau^-$ close to the $Z$ mass \cite{Heister:2002ik} (our couplings can be expressed in terms of theirs as $(d_\tau^Z,a_\tau^Z)=(2m_\tau\sin\theta_W\cos\theta_W)\times(d_\tau^W,\mu_\tau)$.)  At the 95\% c. l. they quote 
\begin{eqnarray}
Re(d_\tau^W) < 5.0 \times 10^{-18}{\rm ~e~cm}, &{\rm ~or~}&m_\tau |d_\tau^Z| \lsim 6.7\times 10^{-4} \nonumber \\ 
Re(\mu_\tau) < 1.1 \times 10^{-3}, &{\rm ~or~}&m_\tau |a_\tau^Z| \lsim 1.6\times 10^{-3}
\end{eqnarray}
A proposal to improve on the result for $a_\tau^\gamma$ by an order of magnitude  with high luminosity B factories exists \cite{Fael:2013ij}. 

With the aid of Eq.~\ref{fitg} we can find the region of parameter space in the $\tau$-gluonic couplings that is allowed at $1\sigma$ by requiring the total di-tau cross-section to deviate from its SM value (equivalently from the di-muon cross-section) by at most $14\%$, we find,
\begin{eqnarray}
\left(|d_{\tau G}|^2+|d_{\tau \tilde{G}}|^2\right) \lsim 0.93.
\label{dtgbound}
\end{eqnarray}
Note that these bounds are about five times better than those for $d_{\tau W}$ for $\Lambda=1$~TeV, corroborating the impact of the much higher gluon parton luminosities. The result, Eq.~\ref{dtgbound}, is very close to what was obtained in Ref.~\cite{Potter:2012yv} requiring $3\sigma$ statistical sensitivity with an integrated luminosity of 10~fb$^{-1}$. Of course, since these couplings originate at dimension eight, their impact drops significantly faster than that of the anomalous dipole couplings for higher new physics scales.

It may be interesting to consider the $Z$-resonance region as well given the very large number of events expected there. A simulation for $60<m_{\tau\tau}<120$~GeV is presented in Figure~\ref{dtVZsig14} and the corresponding fit in Eq.~\ref{fitZregV}. In this case the reported systematic error in the measurement of the $Z$ cross-section using the di-tau channel by CMS is $8\%$ \cite{Chatrchyan:2011nv} and it is the main source of error already. Assuming a comparison to the theoretical cross-section at the $10\%$ will be possible at 14~TeV we find constraints only slightly weaker than Eq.~\ref{ditauresults}, by about $30\%$. 

\section{Higgs production in association with a $\tau$-lepton pair}

We now turn our attention to the associated production of a Higgs boson with a $\tau$-lepton pair. The motivation to study this process is simple, gauge invariance of the effective Lagrangian relates the dipole couplings of the $\tau$-lepton to couplings involving the Higgs boson. At the same time a similar study of top-quark dipole couplings in Ref.~\cite{Hayreter:2013kba} suggests that the two processes can yield comparable constraints. From the experimental perspective, there are Higgs boson searches in the $HV$, $V=Z,W$ modes underway \cite{Chatrchyan:2012qr,Aad:2012gxa}. Although at the moment they follow a different analysis, they can in principle also be conducted for the case $Z\to \tau^+\tau^-$. 

\begin{figure}[thb]
\includegraphics[width=0.45\textwidth]{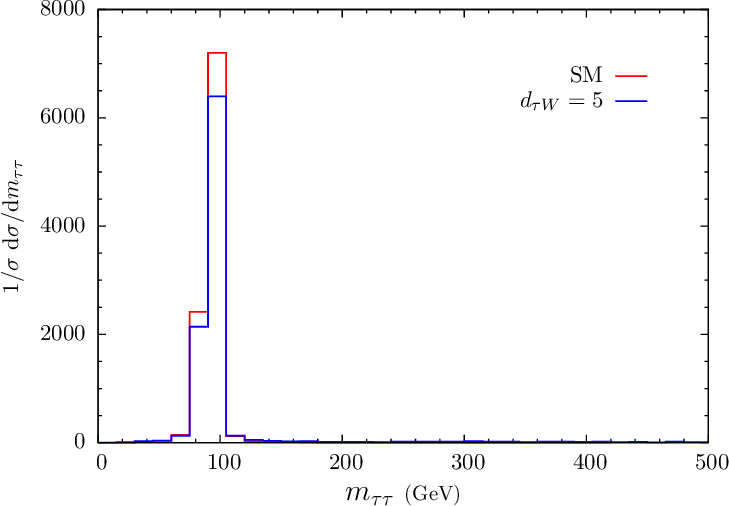} \hspace{1.0cm}
\includegraphics[width=0.45\textwidth]{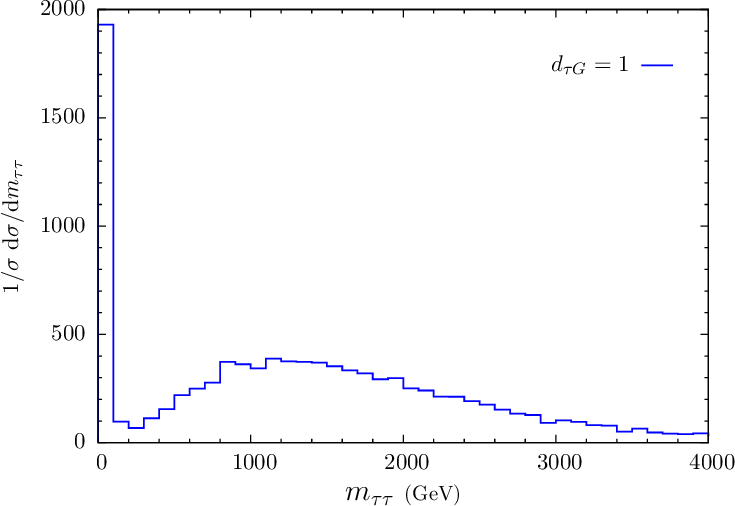}
\caption{Differential decay distribution $d\sigma(pp\to \tau^+\tau^- h)/dm_{\tau\tau}$ for: left plot the SM and for $d_{\tau W}=5$ and; right plot $d_{\tau G}=1$.}
\label{mtautau}
\end{figure}

To estimate the event rates that can be expected we take the SM prediction for $m_h=125$~GeV at 14~TeV from the CERN Report 2011-002  \cite{ppZh} $\sigma(pp\to Zh)=(0.8830^{+6.4\%}_{-5.5\%})$~pb, and multiply it by $Br(Z\to \tau^+\tau^-)$ to  find  $\sigma(pp\to Zh \to \tau^+\tau^- h)=(29.8^{+1.9}_{-1.6})$~fb. In Figure~\ref{mtautau} we show the distributions $d\sigma(pp\to \tau^+\tau^- h)/dm_{\tau\tau}$ for the SM, for $d_{\tau W}=5$, and for the case $d_{\tau G}=1$ which correspond to what can be constrained in Drell-Yan according to Figure~\ref{contourhigh} and to Eq.~\ref{dtgbound}. The distributions show that whereas the SM is sharply peaked at the $Z$-resonance as is the $d_{\tau W}$ induced rate, the $d_{\tau G}$ case is not. Of course the contribution to the rate  beyond the scale $\Lambda$ (which we have chosen to be 1~TeV for this study) is not physical and simply reflects the nature of the effective theory. This could be addressed by rescaling the bounds on the anomalous couplings by increasing 
the new physics scale if desired.

\begin{figure}[thb]
\includegraphics[width=0.45\textwidth]{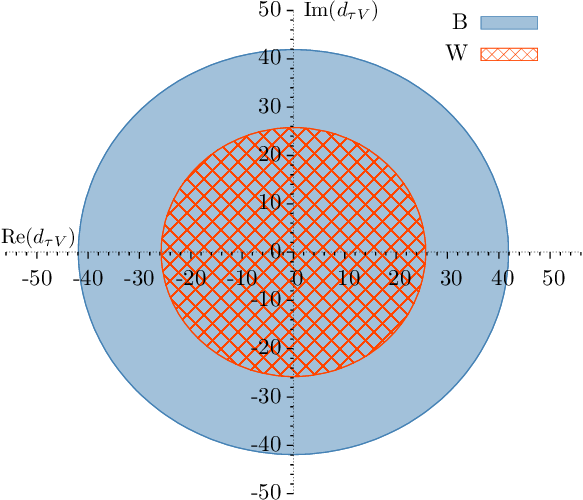} \hspace{0.8cm}
\includegraphics[width=0.48\textwidth]{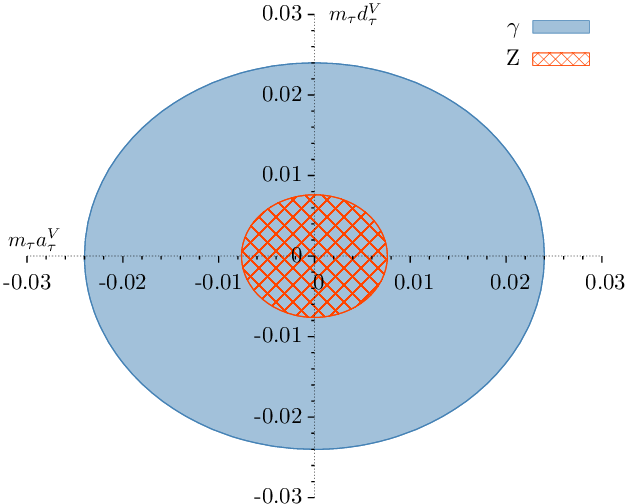}
\caption{Parameter space allowed at $1\sigma$ by requiring the total cross-section $pp\to \tau^+\tau^-h$ at 14~TeV with $60<m_{\tau\tau}<120$~GeV to deviate from its SM value by at most a factor of 2 (or remain below 26~fb).}
\label{contourtthZ}
\end{figure}

Since the contribution of $d_{\tau W}$ is largest at the $Z$ resonance we first consider its constraints from $60<m_{\tau\tau}<120$~GeV. The numerical cross-sections obtained for sample values of the anomalous couplings, along with the corresponding fit are displayed in the right Figure~\ref{dtVZsig14} and in the second Eq.~\ref{fitZregV}. If the cross-section can be constrained to be at most twice the SM value (or below 26~fb) one would obtain the bounds shown in Figure~\ref{contourtthZ}.

\begin{figure}[thb]
\includegraphics[width=0.45\textwidth]{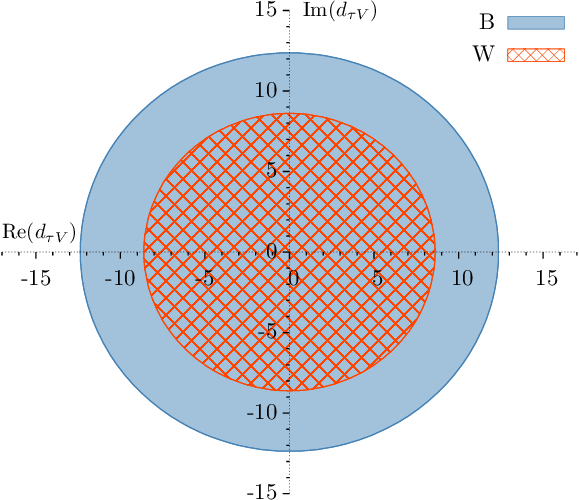} \hspace{0.8cm}
\includegraphics[width=0.48\textwidth]{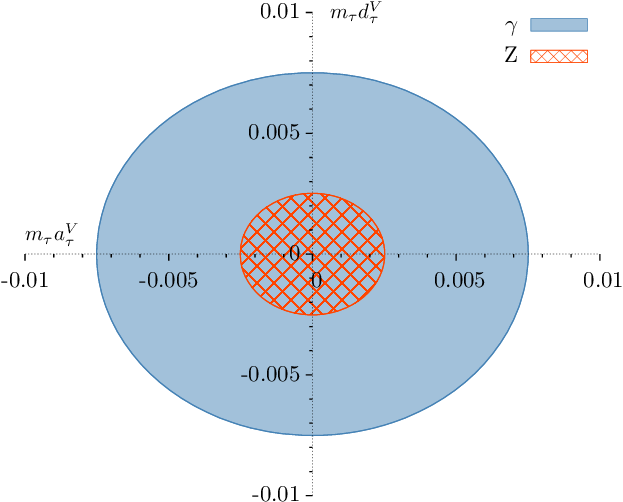}
\caption{Parameter space allowed at $1\sigma$ by requiring the total cross-section $pp\to \tau^+\tau^-h$ at 14~TeV with $m_{\tau\tau}>120$~GeV to be at most 5~fb (50 times the SM value).}
\label{contourtth}
\end{figure}

In view of Figure~\ref{mtautau} we limit our study for the $d_{\tau G}$ coupling to the high di-tau invariant mass region, $m_{\tau\tau}>120$~GeV, where the SM background is smallest. The numerical cross-sections obtained for sample values of the anomalous couplings, along with the corresponding fit are displayed in the right Figure~\ref{dtvsig14} and in the second Eq.~\ref{fitV} respectively for anomalous dipole couplings and in Figure~\ref{dtgsig14} and in the second Eq.~\ref{fitg} for the gluon-leptonic couplings. Assuming that $\sigma(pp\to \tau^+\tau^-h)$ with $m_{\tau\tau}>120$~GeV is measured to within a factor of fifty of its SM value (that is, the cross-section can be bounded by 5~fb) yields an allowed region for the anomalous dipole couplings displayed in Figure~\ref{contourtth}.
The bounds from Figure~\ref{contourtth} are comparable to those in Figure~\ref{contourhigh}, indicating that a 14\% measurement of the Drell-Yan di-tau cross-section yields similar bounds to a constraint $\sigma(pp\to \tau^+\tau^-h) \lsim 5$~fb in the region  $m_{\tau\tau}>120$~GeV. 

The situation is different for the lepton-gluonic couplings where Eq.~\ref{fitg} implies that an experimental constraint on $\sigma(pp\to \tau^+\tau^-h)$ at the $50$~fb level (500 times larger than the SM) is enough to match the constraints on the tau-gluonic couplings from the Drell-Yan process, Eq.~\ref{dtgbound}.
\begin{eqnarray}
\left(|d_{\tau G}|^2+|d_{\tau \tilde{G}}|^2\right) \lsim 20\times {\rm ~experimental~upper~bound~on~}\frac{\sigma(pp\to \tau^+\tau^-h)}{1\rm ~pb}.
\label{dtgboundh}
\end{eqnarray}

\section{Analysis at the $\tau$ decay level}

The constraints that we have obtained in the previous section rely on a measurement of the di-tau cross-section and on the level at which background can be controlled. In this section we explore this in somewhat more detail by including the $\tau$-lepton decay in our analysis as well as the physics backgrounds that exist at that level following the guidelines used in Ref.~\cite{Gupta:2011vt}. We will consider only two cases. First the dilepton mode in which both $\tau$-leptons undergo leptonic decay into muons or electrons:  $pp\to \tau^+\tau^- \to \ell^+ \ell^{\prime -} \slashed{E}_T$, $\ell =\mu,e$. The background for this mode includes
\begin{itemize}
\item $t\bar{t}$ pairs in which both $b$-jets are missed. We model this by requiring $p_{Tb}<20$~GeV.
\item $ZZ$ pairs in which one $Z$ decays to charged leptons and the other one to neutrinos. This background can be suppressed considerably by requiring the di-lepton mass to be above 120~GeV.
\item $W^+ W^-$ pairs.
\item For same flavor pairs we also have direct Drell-Yan production of $\ell^+\ell^-$. This background can be eliminated with the requirement that the events have a minimum missing $E_T$ of 10~GeV.
\end{itemize}
Second we consider modes in which one $\tau$-lepton decays leptonically to a muon or an electron and the other one hadronically. The background for these modes consists of the $t\bar{t}$, $W^+ W^-$, or $ZZ$ as in the previous modes and in addition the more important of $W$ plus one jet ($Wj$) production. We will limit our analysis to the two hadronic decay modes $\tau^\pm \to \pi^\pm \nu$ and $\tau^\pm \to \rho^\pm\nu$ which together account for a branching ratio near 36\%.  For the $Wj$ background that meets our selection cuts, we assume a probability of 0.3\% that the QCD jet will fake a $\tau$ jet and reduce these events accordingly. This number is taken from studies by ATLAS and CMS \cite{fakejet}. 

\begin{figure}[thb]
\includegraphics[width=0.335\textwidth]{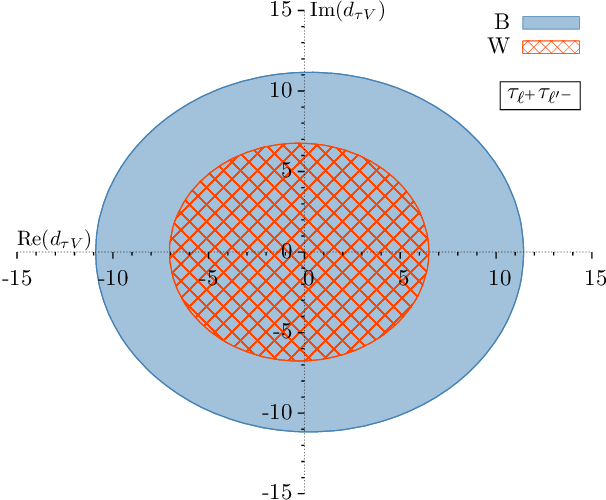}
\includegraphics[width=0.32\textwidth]{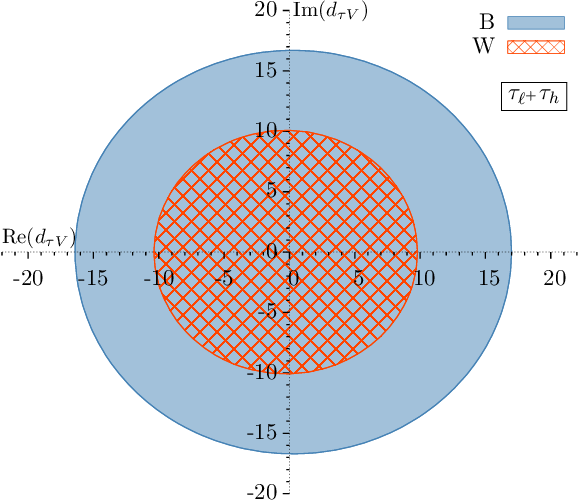}
\includegraphics[width=0.32\textwidth]{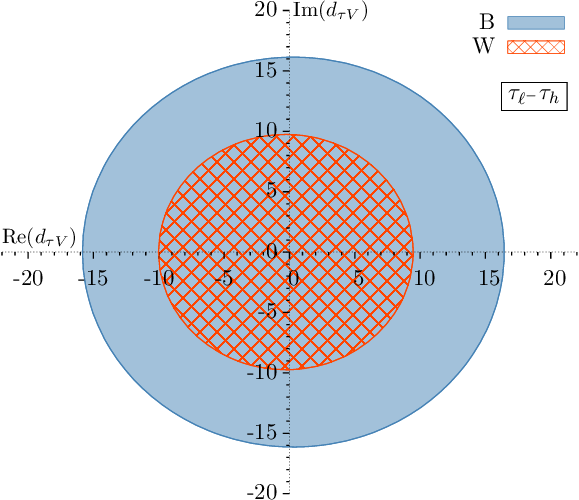}
\caption{Limits that can be obtained on the anomalous dipole couplings of the $\tau$-lepton assuming a comparison between theory and experiment at the 14\% level in the modes $\tau_{\ell^+} \tau_{\ell^{\prime -}}$ (left), $\tau_{\ell^+}\tau_h$ (center) and  $\tau_{\ell^-}\tau_h$ (right).}
\label{taudecaylimit}
\end{figure}

In all cases we generate the signal and background events using  {\tt MadGraph5} and implement the $\tau$-lepton decay modes with the package {\tt DECAY} \cite{decay}.  We use the following basic cuts on the 
leptons and the jets: $p_T > 6$ GeV, $|\eta| < 2.5$ and $\Delta R_{ik} > 0.4$, where the indices $i, k = \ell, j$. In addition we require the high invariant mass region with the cuts $m_{\ell\ell^\prime}>120$ GeV or $m_{\ell j}>120$ GeV. We suppress the backgrounds with the cuts detailed above. The results for several values of the anomalous dipole couplings, as well as the corresponding fits are presented in the Appendix in Figure~\ref{taudecayf} and Eq.~\ref{fitstaudecay}. The latter show two numbers for the case of vanishing anomalous couplings: the first number is the remaining background after the cuts and the second number is the SM contribution to the process. If we assume again that comparison between theory and experiment will be limited by the 14\% systematic error mentioned before, we obtain allowed regions shown in Figure~\ref{taudecaylimit}.
We see that the limit that can be obtained with di-lepton modes is comparable to that one quoted before at the di-tau level. The limits in the $\tau_\ell \tau_h$ modes are about a factor of two worse.

\section{Summary}

We have examined the possible limits that can be placed on certain anomalous couplings of $\tau$-leptons at the LHC. We have considered the four dipole-type couplings that appear at dimension six in the effective Lagrangian as well as the two $\tau$-gluon couplings that appear at dimension eight. We have found the constraints that can be placed from a measurement of the high invariant mass region $m_{\tau\tau}>120$~GeV Drell-Yan di-tau cross-section. We have assumed that the comparison between theory and experiment will be limited by the current 14\% maximum error in background estimation for high invariant mass di-tau pairs in CMS. Under this assumption, for the dipole-type couplings we find limits that are comparable to the best existing limit on $(g-2)$ for the $\tau$, within factors of two of the best limit for the $\tau$-edm and 5-10 times worse than the Belle limits on the dipole couplings of the $\tau$ to the $Z$. For new physics at the TeV scale we find that comparable or slightly better constraints 
can be placed on the dimension eight $\tau$-gluonic couplings.

Using the fully gauge invariant effective Lagrangian we related these couplings to new contributions to the associated production of a $\tau$-lepton pair and a Higgs boson. The cross-section for this process is very small, of order 0.1~fb. We find that in order to extract constraints on the dipole couplings that are competitive with those from Drell-Yan one needs to bound $\sigma(pp\to \tau^+\tau^-h)$ at the 5 fb level. On the other hand one only needs to bound it at the 50~fb level (or 500 times the SM) to constrain the $\tau$-gluonic couplings at the same level as with a 14\% measurement of the Drell-Yan cross-section.

Finally we have repeated our study of the Drell-Yan di-tau cross-section allowing for  di-tau decay into di-leptons (muons or electrons) or into lepton plus jet channels keeping the $\pi\nu$ and $\rho\nu$ modes. We find that there is not much loss in sensitivity, mostly because the limitation of the measurement had already been assumed to be due to systematics.

\begin{acknowledgments}

This work was supported in part by the DOE under contract number DOE under contract number DE-SC0009974. We are grateful to David Atwood and Chunhui Chen for useful discussions.

\end{acknowledgments}

\appendix

\section{Events with $\tau$-leptons as final states}

In this appendix we present our numerical results. We begin our calculation by implementing the Lagrangian of Eq.~\ref{ginvedm} and Eq.~\ref{taugluon} into {\tt MadGraph5} with the aid of {\tt FeynRules} \cite{Christensen:2008py}. We use the resulting UFO model file to generate $\tau$-lepton pair events as well as $\tau^+\tau^-h$ events for several values of $d_{\tau W}$, $d_{\tau B}$, $d_{\tau G}$ and $d_{\tau \tilde{G}}$ in a range that changes the $\tau^+\tau^-$ SM cross-section by factors of a few. We then fit  the numerical results to obtain approximate expressions for cross-sections in terms of  the new couplings. We repeat this procedure for two different invariant mass regions. 

\subsection{$\tau$-lepton pairs at the high invariant mass region}

\begin{figure}[thb]
\includegraphics[width=0.46\textwidth]{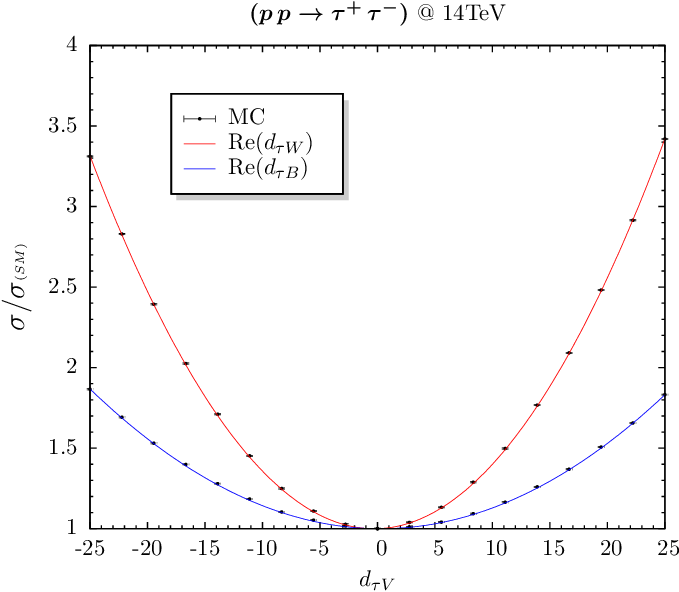} \hspace{1.0cm}
\includegraphics[width=0.45\textwidth]{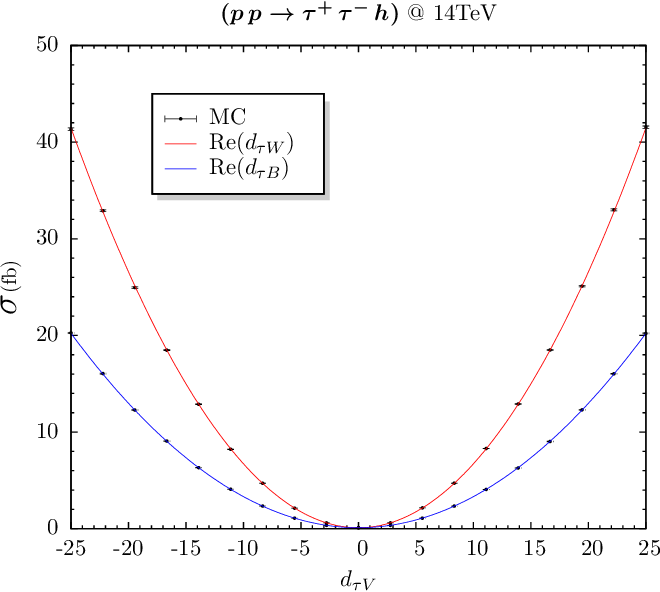}
\caption{Cross-section for the processes $pp\to \tau^+\tau^-$ (left) and $pp\to \tau^+\tau^-h$ (right) at 14 TeV as calculated with {\tt MadGraph5} for different values of the anomalous coupling $d_{\tau V}$ with $m_{\tau\tau}>120$~GeV and the corresponding fits.}
\label{dtvsig14}
\end{figure}

Our first study is for the high energy region defined by $m_{\tau\tau}>120$~GeV and the results are shown in Figure~\ref{dtvsig14}.
We fit the MonteCarlo (MC) points to a quadratic equation in the anomalous couplings since they occur linearly in the amplitudes. The result of this fit, normalized to the SM for the case of $\tau$-lepton pair production (which is calculated by  {\tt MadGraph5} to be $\sigma(pp\to \tau^+\tau^-)_{SM}= (9.03\pm 0.02) {\rm ~pb}$), is
\begin{eqnarray}
\frac{\sigma(pp\to \tau^+\tau^-)}{\sigma_{SM}} &=& 1+2.1 \times 10^{-3} {\rm Re}(d_{\tau W})+3.8 \times 10^{-3}  |d_{\tau W}|^2  \nonumber \\
&-& 7.2\times 10^{-4} {\rm Re}(d_{\tau B})+1.4 \times 10^{-3}  |d_{\tau B}|^2 \nonumber \\
\sigma(pp\to \tau^+\tau^-h) &=& \left(0.1 + 6.6\times 10^{-2}|d_{\tau W}|^2+ 3.2\times 10^{-2}|d_{\tau B}|^2\right)~{\rm ~fb}
\label{fitV}
\end{eqnarray}
The last cross-section is to be compared to the very low SM cross-section 
$\sigma(pp\to \tau^+\tau^-h)_{SM}= 9.6\times 10^{-5}{\rm ~pb}$

\begin{figure}[thb]
\includegraphics[width=0.45\textwidth]{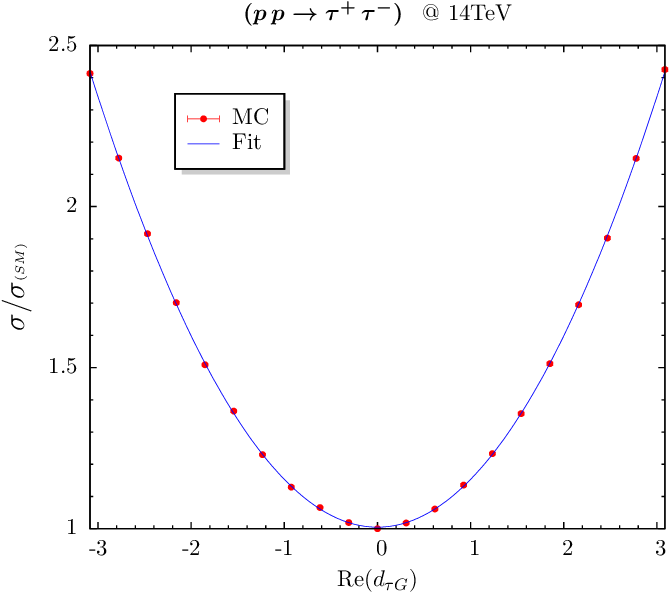} \hspace{1.0cm}
\includegraphics[width=0.45\textwidth]{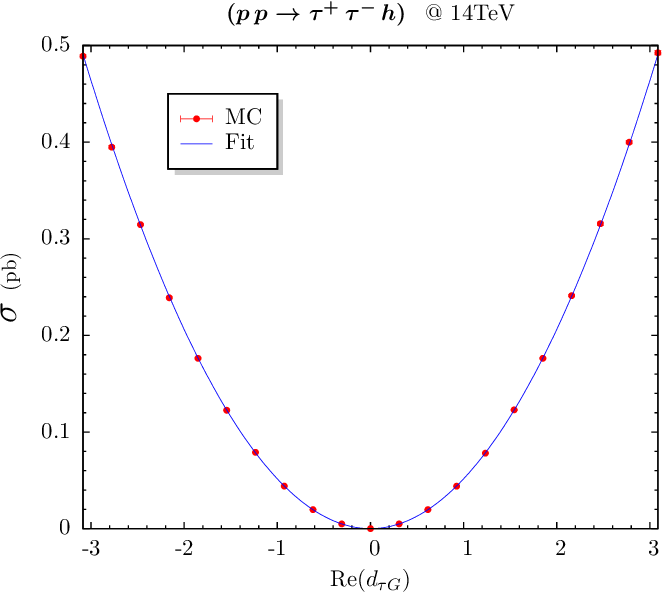}
\caption{Cross-section for the processes $pp\to \tau^+\tau^-$ (left) and $pp\to \tau^+\tau^-h$ (right) at 14 TeV as calculated with {\tt MadGraph5} for different values of the anomalous coupling $d_{\tau G}$  with $m_{\tau\tau}>120$~GeV  and the corresponding fits.}
\label{dtgsig14}
\end{figure}

Our results for the tau-gluon couplings are shown in Figure~\ref{dtgsig14}.
As expected from the results of Ref.~\cite{Potter:2012yv}, the effect on the cross-section from Re$(d_{\tau G})$, Re$(d_{\tau \tilde{G}})$, Im$(d_{\tau G})$ and  Im$(d_{\tau \tilde{G}})$ is the same so we show only one of them in the figure. The corresponding (quadratic again) fit is given by
\begin{eqnarray}
\frac{\sigma(pp\to \tau^+\tau^-)}{\sigma_{SM}} &=& 1+0.15 \left(|d_{\tau G}|^2+|d_{\tau \tilde{G}}|^2\right)\nonumber \\
\sigma(pp\to \tau^+\tau^-h) &=& \left(1\times10^{-4}+5.15\times 10^{-2} \left(|d_{\tau G}|^2+|d_{\tau \tilde{G}}|^2\right)\right)~{\rm ~pb}
\label{fitg}
\end{eqnarray}

We generated $10^4$ events for each of our sample coupling values and the errors in the cross-sections are the statistical error calculated by {\tt MadGraph5}. For the $\tau^+\tau^-$ samples $0.02$~pb and for the $\tau^+\tau^-h$ samples
$4\times 10^{-4}$~fb.

\subsection{$\tau$-lepton pairs near the $Z$-resonance region}

\begin{figure}[thb]
\includegraphics[width=0.46\textwidth]{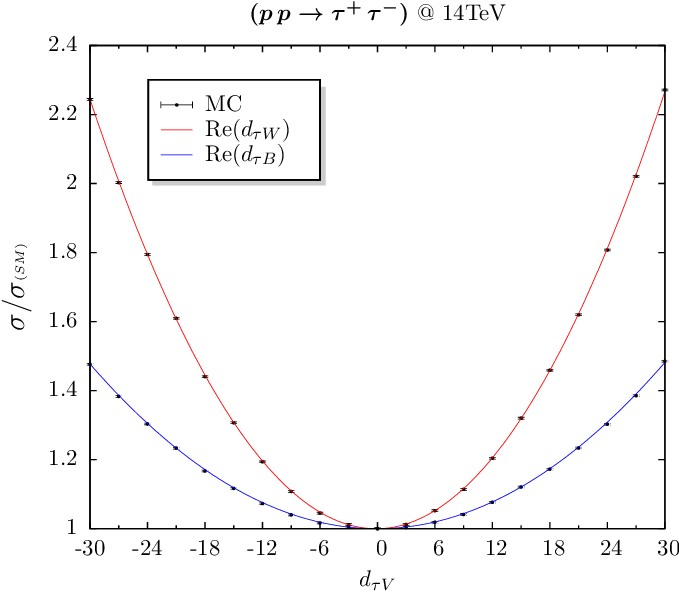} \hspace{1.0cm}
\includegraphics[width=0.45\textwidth]{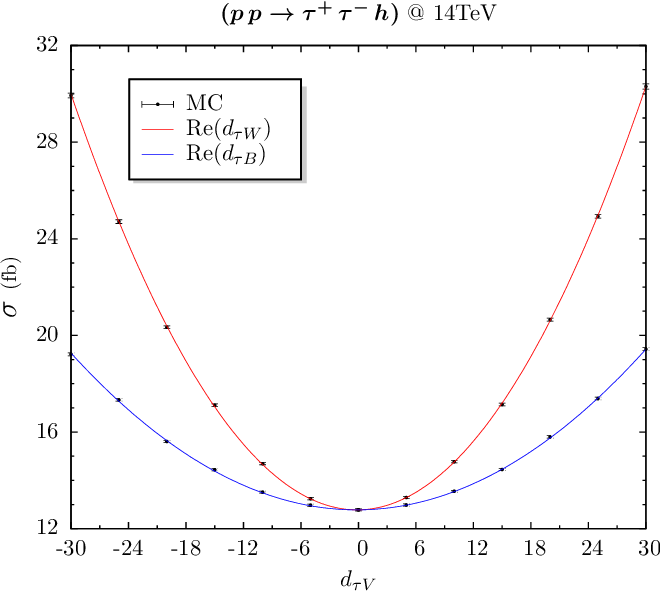}
\caption{Cross-section for the processes $pp\to \tau^+\tau^-$ (left) and $pp\to \tau^+\tau^-h$ (right) with $60< m_{\tau\tau} < 120$~GeV at 14 TeV as calculated with {\tt MadGraph5} for different values of the anomalous coupling $d_{\tau V}$ and the corresponding fits.}
\label{dtVZsig14}
\end{figure}

We next consider the $Z$ resonance region, defined by $60< m_{\tau\tau} < 120$~GeV. Our results for this case are shown in Figure~\ref{dtVZsig14}

The SM cross-section in this case is calculated by {\tt MadGraph5} to be $\sigma(pp\to \tau^+\tau^-)_{SM} = (692.1 \pm 1.3)~{\rm ~pb}$ and $\sigma(pp\to \tau^+\tau^- h)_{SM} = (12.78 \pm 0.04)~{\rm ~fb}$ and the fit to the new physics (NP) is given by
\begin{eqnarray}
\frac{\sigma(pp\to \tau^+\tau^-)}{\sigma_{SM}} &=& 1+3.78 \times 10^{-4}\ {\rm Re}(d_{\tau W})+1.4\times 10^{-3}\  |d_{\tau W}|^2  \nonumber \\
&+& 7.64\times 10^{-5}\  {\rm Re}(d_{\tau B})+5.33 \times 10^{-4}\  |d_{\tau B}|^2 \nonumber \\
\sigma(pp\to \tau^+\tau^-h) &=& \left(12.78 + 1.9\times 10^{-2} \ |d_{\tau W}|^2 + 7.3\times 10^{-3}\ |d_{\tau B}|^2\right){\rm ~fb}
\label{fitZregV}
\end{eqnarray}

\subsection{$\tau_\ell \tau_{\ell^\prime}$ and $\tau_\ell \tau_h$ Modes}

\begin{figure}[thb]
\includegraphics[width=0.45\textwidth]{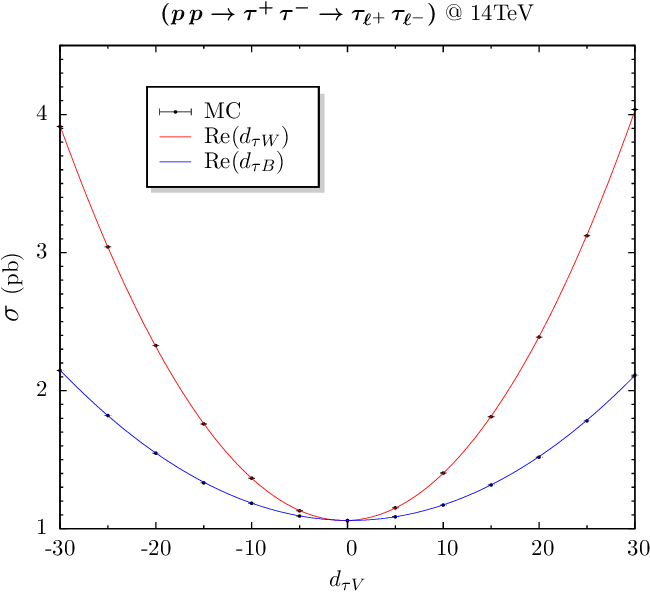} \hspace{1.0cm}
\includegraphics[width=0.45\textwidth]{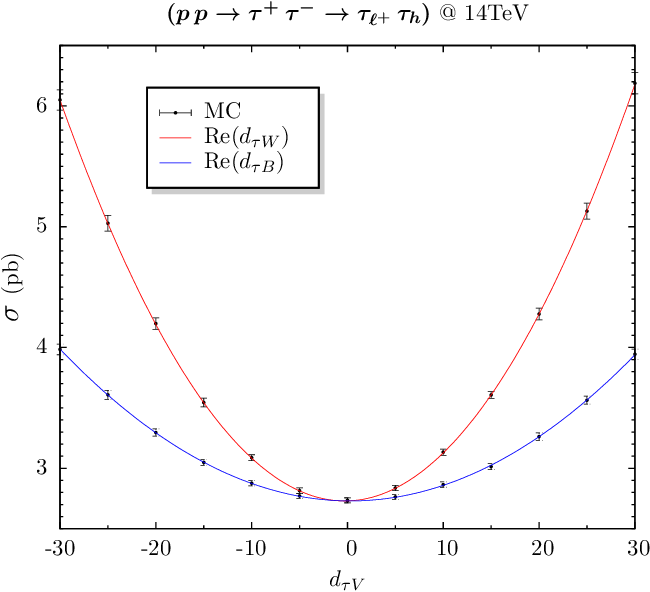}
\includegraphics[width=0.45\textwidth]{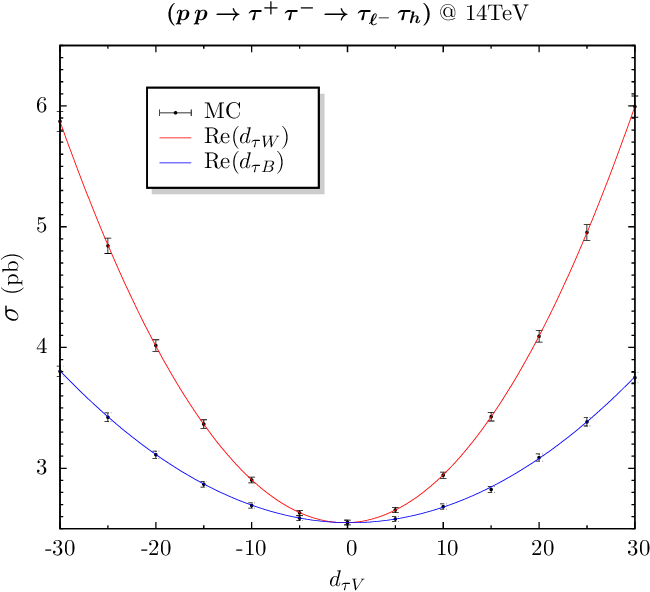}
\caption{Cross-section for the processes $pp\to \tau_{\ell^+}\tau_{\ell^{\prime-}}$, $pp\to \tau_{\ell^+}\tau_h$ and $pp\to \tau_{\ell^-}\tau_h$ with cuts described in the text at 14 TeV as calculated with {\tt MadGraph5} and {\tt DECAY}  for different values of the anomalous coupling $d_{\tau V}$ and the corresponding fits.}
\label{taudecayf}
\end{figure}

We perform a partial study of what happens at the tau decay level using the $\tau_\ell \tau_{\ell^\prime}$ and $\tau_\ell \tau_h$ channels. After starting with $\tau$-lepton pair samples, we allow the $\tau$-leptons to decay either leptonically to a muon or electron plus two neutrinos or into a $\tau$-jet. For the hadronic decay we only include the $\pi\nu$ and $\rho\nu$ modes. All these decays are performed with the {\tt DECAY} package. The results are shown in Figure~\ref{taudecayf}.
We do not show figures for the imaginary parts of the couplings as they are indistinguishable from these ones. 
The corresponding fits are given by (the statistical error computed by  {\tt MadGraph5}  is 0.032~pb for di-lepton case)
\begin{eqnarray}
\sigma(pp\to \tau_{\ell^+}\tau_{\ell^{\prime-}}) &=& 0.27{\rm~pb~} + \left(0.79 + 1.80\times10^{-3} {\rm~Re}(d_{\tau W}) + 3.24\times10^{-3} \left|d_{\tau W}\right|^2\right. \nonumber \\
                    &-& \left. 6.44\times10^{-4} {\rm~Re}(d_{\tau B}) + 1.19\times10^{-3} \left|d_{\tau B}\right|^2\right){\rm~pb~} \nonumber \\
\sigma(pp\to \tau_{\ell^+}\tau_h) &=& 1.80 {\rm~pb~} + \left(
0.95 + 2.15\times10^{-3} {\rm~Re}(d_{\tau W}) + 3.76\times10^{-3}  \left|d_{\tau W}\right|^2 \right. \nonumber \\
                   &-&\left. 8.07\times10^{-4} {\rm~Re}(d_{\tau B}) + 1.37\times10^{-3} \left|d_{\tau B}\right|^2\right){\rm~pb~} \nonumber \\
\sigma(pp\to \tau_{\ell^-}\tau_h) &=& 1.60 {\rm~pb~} + \left(
0.93 +  2.07\times10^{-3} {\rm~Re}(d_{\tau W}) + 3.76\times10^{-3}  \left|d_{\tau W}\right|^2\right. \nonumber \\
                   &-&\left. 8.22\times10^{-4} {\rm~Re}(d_{\tau B}) + 1.37\times10^{-3} \left|d_{\tau B}\right|^2\right){\rm~pb~}
\label{fitstaudecay}
\end{eqnarray}
In these equations we have separated the contribution from the background as described in the main section (first number) and the signal.

\end{document}